\documentclass[runningheads]{llncs}
\usepackage{graphicx}
\usepackage[english]{babel}
\usepackage[utf8]{inputenc}
\usepackage[T1]{fontenc}
\usepackage[export]{adjustbox}
\usepackage{booktabs}
\usepackage{ae,aecompl}

\usepackage{pgfplots}
\pgfplotsset{compat=1.12}
\usepackage[group-separator={,}]{siunitx}
\usepackage{amsmath}
\usepackage{wrapfig}
\usepackage{algpseudocode}
\usepackage{graphicx}
\usepackage{algorithm}
\usepackage{breqn}
\usepackage{algpseudocode}
\usepackage{float}
\usepackage{tikz}
\usepackage{pifont}
\usepackage{eufrak}
\usepackage{caption}
\usepackage{listings}
\usepackage[toc,page]{appendix}
\usepackage{color}
\definecolor{lightgray}{rgb}{.9,.9,.9}
\definecolor{darkgray}{rgb}{.4,.4,.4}
\definecolor{purple}{rgb}{0.65, 0.12, 0.82}

\def\bitcoinA{%
	\leavevmode
	\vtop{\offinterlineskip 
		\setbox0=\hbox{B}%
		\setbox2=\hbox to\wd0{\hfil\hskip-.03em
			\vrule height .3ex width .15ex\hskip .08em
			\vrule height .3ex width .15ex\hfil}
		\vbox{\copy2\box0}\box2}}

\begin{document}
\title{Topological Analysis of Bitcoin's Lightning Network}
%
%
\author{István András Seres\inst{1}\orcidID{0000-0003-0143-4057} \and
László Gulyás\inst{1} \and
Dániel A. Nagy\inst{1} \and
Péter Burcsi\inst{1}\orcidID{0000-0003-3306-6500}}
\authorrunning{I.A. Seres et al.}
%
\institute{Eötvös Loránd University, Budapest 1053, HUN
\\
\email{istvanseres@caesar.elte.hu,gulyahps.elte.hu,nagy.da@gmail.com,bupe@inf.elte.hu}}
\maketitle              
\begin{abstract}
Bitcoin's Lightning Network (LN) is a scalability solution for Bitcoin allowing transactions to be issued with negligible fees and settled instantly at scale. In order to use LN, funds need to be locked in payment channels on the Bitcoin blockchain (Layer-1) for subsequent use in LN (Layer-2). LN is comprised of many payment channels forming a payment channel network. LN's promise is that relatively few payment channels already enable anyone to efficiently, securely and privately route payments across the whole network. In this paper, we quantify the structural properties of LN and argue that LN's current topological properties can be ameliorated in order to improve the security of LN, enabling it to reach its true potential.

\end{abstract}
Since its launch, Bitcoin~\cite{nakamoto2008bitcoin} gained a huge popularity due to its publicly verifiable, decentralized, permissionless and censorship-resistant nature. This tremendous popularity and increasing interest in Bitcoin pushed its network's throughput to its limits. Without further advancements, the Bitcoin network can only settle $7$ transactions per second (tps), while mainstream centralized payment providers such as Visa and Mastercard can process approximately \num[group-separator={,}]{40000} tps in peak hours. Moreover one might need to pay large transaction fees on the Bitcoin network, while also need to wait 6 new blocks ($\sim 1$ hour) to be published in order to be certain enough that the transaction is included in the blockchain.

To alleviate these scalability issues the Lightning Network (LN) was designed in 2016 \cite{poon2016bitcoin}, and launched in 2018, January. The main insight of LN is that transactions can be issued off-blockchain in a trust-minimized manner achieving instant transaction confirmation times with negligible fees, whilst retaining the security of the underlying blockchain. Bidirectional payment channels can be formed on-chain using a construction called Hashed Timelock Contracts (HTLC). Later several payments can take place in a payment channel. The main advantage of payment channels is that one can send and receive thousands of payments with essentially only 2 on-chain transations: the opening and closing channel transactions. 

Using these payment channels as building blocks one might establish a payment channel network, where it is not necessary to have direct payment channels between nodes to transact with each other, but they could simply route their payments through other nodes' payment channels.
Such a network can be built, because LN achieves payments to be made without any counterparty risk, however efficient and privacy-preserving payment routing remains a challenging algorithmic task~\cite{roos2017settling}. 

\textbf{Our contributions.} We empirically measure\footnote{https://github.com/seresistvanandras/LNTopology} and describe LN's topology and show how robust it is against both random failures and targeted attacks. These findings suggest that LN's topology can be ameliorated in order to achieve its true potential.
\section{Background on Lightning Network}
In this section we provide a short recap on how LN works. In the following we will use the terms Layer-2 and off-chain interchangeably. LN is a so-called Layer-2 technology, which allows participants issuing transactions without sending a Layer-1 transaction on the Bitcoin parent chain. All parties cooperatively
open a channel by locking collateral on the blockchain. The funds can only be released by unanimous agreement or through a pre-defined refund condition~\cite{cryptoeprint:2019:360}. One of the greatest challenge of Layer-2 technologies is to solve how participants can agree on new state updates in a trustless or trust-minimized manner.

Let's take the following toy example: Alice and Bob creates by a single Layer-1 transaction a payment channel with initial balances $10\bitcoinA$ and $0\bitcoinA$ respectively. Straightaway Alice can issue off-chain transactions to Bob up to $10\bitcoinA$. Let's assume Alice issued $3$ off-chain transactions to Bob each worth of $1\bitcoinA$. Afterwards Alice's and Bob's balance should be $7\bitcoinA$ and $3\bitcoinA$ and neither Alice, nor Bob should be able to redeem previous balances on the parent chain. LN achieves this by a replace by revocation mechanism, namely both parties collectively
authorize a new state before revoking the previous state.
Upon dispute, the blockchain provides a time period for
parties to prove that the published state is a revoked state \cite{cryptoeprint:2019:360}. Revoking old channel states is achieved by the exchange of revocation secrets. These secrets, hash preimages, are needed to be retained during the channel's lifetime. A penalty mechanism
discourages parties from broadcasting older states. If one party
broadcasts a revoked state, the blockchain accepts within a
time-window proofs of maleficence from the other party. A
successful dispute allots the winning party \textit{all} coins of the
channel. In our example if Alice broadcasts a revoked channel state with balances $8\bitcoinA$ and $2\bitcoinA$, Bob can prove, that Alice maliciously broadcasted a revoked state. The penalty mechanism grants all the $10\bitcoinA$ to Bob.

The great insight of LN, is that if now Alice would like to issue a payment to Cecily, who eventually has already established a payment channel to Bob, then Alice does not need to open a payment channel and create a costly on-chain transaction, rather she can route her payment through her payment channel with Bob to Cecily. However the maximum amount of bitcoins Alice can send to Cecily is the minimum of all the individual balances on the payment route from Alice to Cecily. Hashed time-locked contracts (HTLC) enable routed payments to be atomic. For a technical description of HTLCs and multi-hop payments the astute reader is referred to~\cite{poon2016bitcoin}. 

In the following we will model LN as an undirected, weighted graph, where nodes are entities who can issue payments using payment channels which are the edges of the LN graph. The weight on the edges, capacities, are the sum of individual balances. Note, that in most cases individual balances are not known to outsiders. Only the capacity of a payment channel is public information, however one can effectively assess individual balances with handy algorithms~\cite{cryptoeprint:2019:328}. 

\section{Lightning Network's Topology}\label{lntopology}
\begin{wrapfigure}[21]{r}{6.5cm}
	\begin{tabular}{| l | c |}
		\hline
		Number of nodes & $2344$  \\ \hline
		Number of payment channels & $16617$  \\ \hline
		Average degree & $7.0891$ \\ \hline
		Connected components & $2$  \\ \hline
		Density & $0.00605$ \\ \hline
		Total BTC held in LN & $543.61855$\bitcoinA\\ \hline
		s-metric& $0.6878$ \\ \hline
		Maximal independent set & $1564$ \\ \hline
		Bridges & $530$ \\ \hline
		Diameter & $6$ \\ \hline
		Radius & $3$ \\ \hline
		Mean shortest path &  $2.80623$ \\ \hline
		Transitivity& $0.1046$\\ \hline
		Average clustering coefficient&$0.304$ \\ \hline
		Degree assortativity& $-0.2690$ \\ \hline
	\end{tabular}
	\caption{LN at a glance: basic properties of the LN graph.}\label{fig:properties}
\end{wrapfigure}

LN can be described as a weighted graph $G=(V,E)$, where $V$ is the set of LN nodes and $E$ is the set of bidirectional payment channels between these nodes. We took a snapshot\footnote{https://graph.lndexplorer.com} of LN's topology on the 10th birthday of Bitcoin, 2019 January 3rd. In the following we are going to analyze this dataset. Although the number of nodes and payment channels are permanently changing in LN, we observed through several snapshots that the main topological characteristics (density, average degree, transitivity, nature of degree distribution) remained unchanged. We leave it for future work to analyze the dynamic properties of LN.

LN gradually increased adoption and attraction throughout 2018, which resulted in 3 independent client implementations (c-lightning\footnote{https://github.com/ElementsProject/lightning/}, eclair\footnote{https://github.com/ACINQ/eclair} and lnd\footnote{https://github.com/lightningnetwork/lnd}) and \num[group-separator={,}]{2344} nodes joining LN as of 2019, January 3rd. The density of a graph is defined as $D=\frac{2\mid E\mid}{\mid V\mid \mid V-1\mid}$  which is the ratio of present and potential edges. As it is shown in Figure \ref{fig:properties}. LN is quite a sparse graph. This is further justified by the fact that LN has \num[group-separator={,}]{530} bridges, edges which deletion increases the number of connected components. Although LN consists of $2$ components, the second component has only $3$ nodes. The low transitivity, fraction of present and possible triangles in the graph, highlights the sparseness of LN as well.

\begin{wrapfigure}{r}{0.7\textwidth}	
		\includegraphics[width=0.8\textwidth]{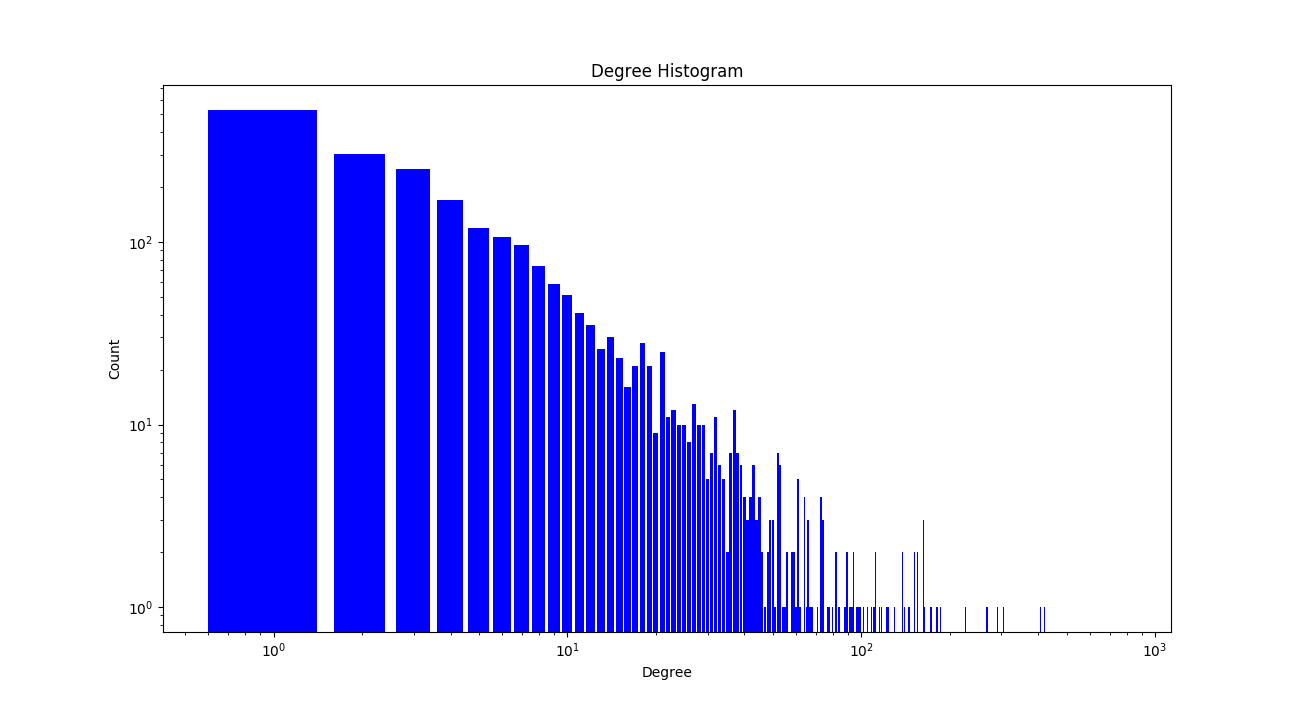}
	\caption{LN's degree distribution}\label{fig:degreedist}	
\end{wrapfigure}

Negative degree assortativity of the graph indicates that on average low degree nodes tend to connect to high degree nodes rather than low degree nodes. Such a dissortative property hints a hub and spoke network structure, which is also reflected in the degree distribution, see Figure~\ref{fig:degreedist}.

Average shortest path length is $2.80623$, without taking into account capacities of edges, which signals that payments can easily be routed with a few hops throughout the whole network. Although this is far from being a straightforward task, since one also needs to take into consideration the capacity of individual payment channels along a candidate path.

When a new node joins LN, it needs to select which other nodes it is trying to connect to. In the \textit{lnd} LN implementation key goals for a node is to optimize its centrality by connecting to central nodes. This phenomena sets up a preferential attachment pattern. Other LN implementations rely on their users to create channels manually, which also most likely leads to users connecting to high-degree nodes. Betweenness centrality of a node $v$ is given by the expression $g(v)=\sum_{s\neq v \neq t}\frac{\sigma_{st}(v)}{\sigma_{st}},$ where $\sigma_{st}$ is the total number of shortest paths between node $s$ and $t$, whilst $\sigma_{st}(v)$ is the number of those paths, that pass through $v$. Closeness centrality  of a node $v$ is defined as $CC(u)=\frac{N}{\sum_{u\neq v}d(u,v)},$ where $N$ is the number of nodes in the graph and $d(u,v)$ is the distance between node $u$ and $v$.  Closeness centrality measures how close a node is to all other nodes. 

Small-world architectures, like LN, exhibit high clustering with short path lenghts. The appropriate graph theoretic tool to asses clustering is the clustering coefficient \cite{watts1998collective}. Local clustering coefficient measures how well a node's neighbors are connected to each other, namely how close they are to being a clique. If a node $u$ has $deg(u)$ neighbors, then between these $deg(u)$ neighbors could be at maximum $\frac{1}{2}deg(u)(deg(u)-1)$ edges. If $N(u)$ denotes the set of $u$'s neighbors, then the local clustering coefficient is defined as $C(u)=\frac{2\mid (v,w): v,w \in N(u) \land (v,w) \in E \mid}{deg(u)(deg(u)-1)}$. 

LN's local clustering coefficient distribution suggestively captures that LN is essentially comprised of a small central clique and a loosely connected periphery.

\subsection{Analysis of LN's degree distribution}

\begin{wrapfigure}{r}{0.7\textwidth}
	\includegraphics[width=0.75\textwidth]{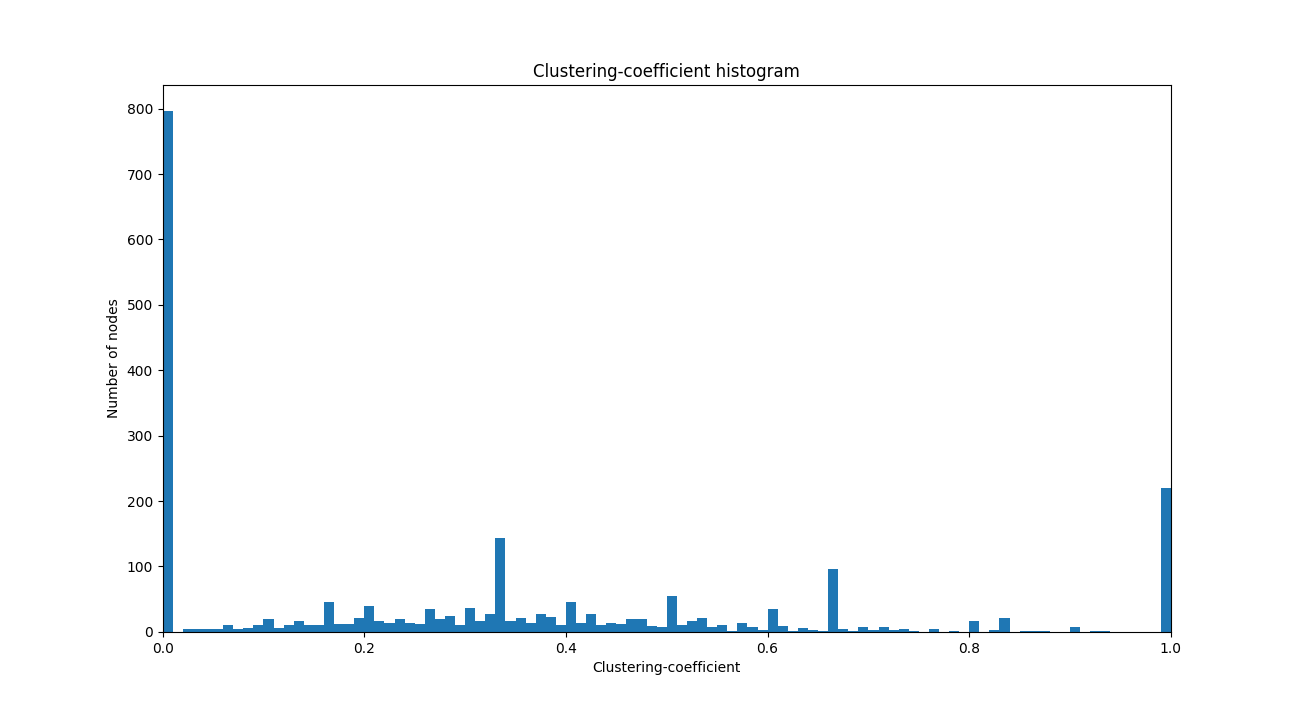}
	\caption{Local clustering coefficient of LN}\label{fig:localclustering}
\end{wrapfigure} 

LN might exhibit scale-free properties as the s-metric suggests. S-metric was first introduced by Lun Li et al. in \cite{li2005towards} and defined as $s(G)=\sum_{(u,v)\in E}deg(u)deg(v)$. The closer to $1$  s-metric of $G$ is, the more scale-free the network. Diameter and radius of LN suggest that LN is a small world. Somewhat scale-freeness is also exhibited in the degree distribution of LN. Majority of nodes have very few payment channels, although there are a few hubs who have significantly more connections as it can be seen in Figure~\ref{fig:degreedist}. The scale-freeness of LN is further justified also by applying the method introduced in \cite{clauset2009power}. The maximum-likelihood fitting asserted that the best fit for the empirical degree distribution is a power law function with exponent $\gamma=-2.1387$. The goodness-of-fit of the scale-free model was ascertained by the Kolmogorov-Smirnov statistic. We found that the $p$-value of the scale-free hypothesis is $p=0.8172$, which is accurate within $0.01$. Therefore the scale-free hypothesis is plausible for the empirical degree distribution of LN.

\section{Robustness of LN}
It is a major question in network science how robust a given network is. LN, just like Bitcoin, is a permissionless network, where nodes can join and leave arbitrarily at any point in time. Nodes can also create new payment channels or close them any time. Furthermore as new payments are made, capacities of payment channels are changing steadily. Despite the dynamic nature of LN, its topology's characteristics remain constant after all.
In this section we investigate how resilient LN is, whether it can effectively withhold random node failures or deliberate attacks.

Measuring robustness means that one gradually removes nodes and/or edges from the graph, until the giant component is broken into several isolated components. The fraction of nodes that need to be removed from a network to break it into multiple isolated connected components is called the percolation threshold and is denoted as $f_c$. In real networks percolation threshold can be well estimated by the point where the giant component first drops below $1\%$ of its original size \cite{barabasi2016network}.  
\subsection{Random Failures}
\begin{wrapfigure}[16]{r}{5cm}
	\begin{tabular}{ | l | c | }
		\hline
		Network & $f_c$  \\ \hline
		Internet & $0.92$  \\ \hline
		WWW & $0.88$ \\ \hline
		US Power Grid & $0.61$  \\ \hline
		Mobil Phone Call & $0.78$ \\ \hline
		Science collaboration& $0.92$ \\ \hline
		E. Coli Metabolism & $0.96$ \\ \hline
		Yeast Protein Interactions & $0.88$ \\ \hline
		LN & $0.96$ \\ \hline
	\end{tabular}
	\caption{Random failures in networks. Values of critical thresholds for other real networks are taken from \cite{barabasi2016network}.}\label{fig:randfailures}
\end{wrapfigure}

Random failures are a realistic attack vector for LN. If nodes happen to be off-line due to bad connections or other reasons, they can not participate in routing payments anymore. Such a failure can be modeled as if a node and its edges are removed from the graph.    

For scale-free networks with degree distribution $P_{k}=k^{-\gamma}$, where $2 < \gamma < 3$ the percolation threshold can be calculated by applying the Molloy-Reed criteria, ie. $f_{c}=1-\frac{1}{\frac{\gamma-2}{3-\gamma}k_{min}^{\gamma-2}k_{max}^{3-\gamma}-1}$, where $k_{min}$ and $k_{max}$ denote the lowest and highest degree nodes respectively. This formula yields $f_c=0.9797$ for LN in case of random failures. This value is indeed close to the percolation threshold measured by network simulation as shown in Figure~\ref{fig:randfailures}, that is, LN provides an evidence of topological stability under random failures. In particular this is due to the fact that in LN a randomly selected node is unlikely to affect the network as a whole, since an average node is not significantly contributing to the network's topological connectivity, see also degree distribution at Figure~\ref{fig:degreedist}.

\subsection{Targeted attacks}
\begin{figure}[h]
	\centering
	\begin{minipage}{.5\textwidth}
		\centering
		\includegraphics[width=\linewidth]{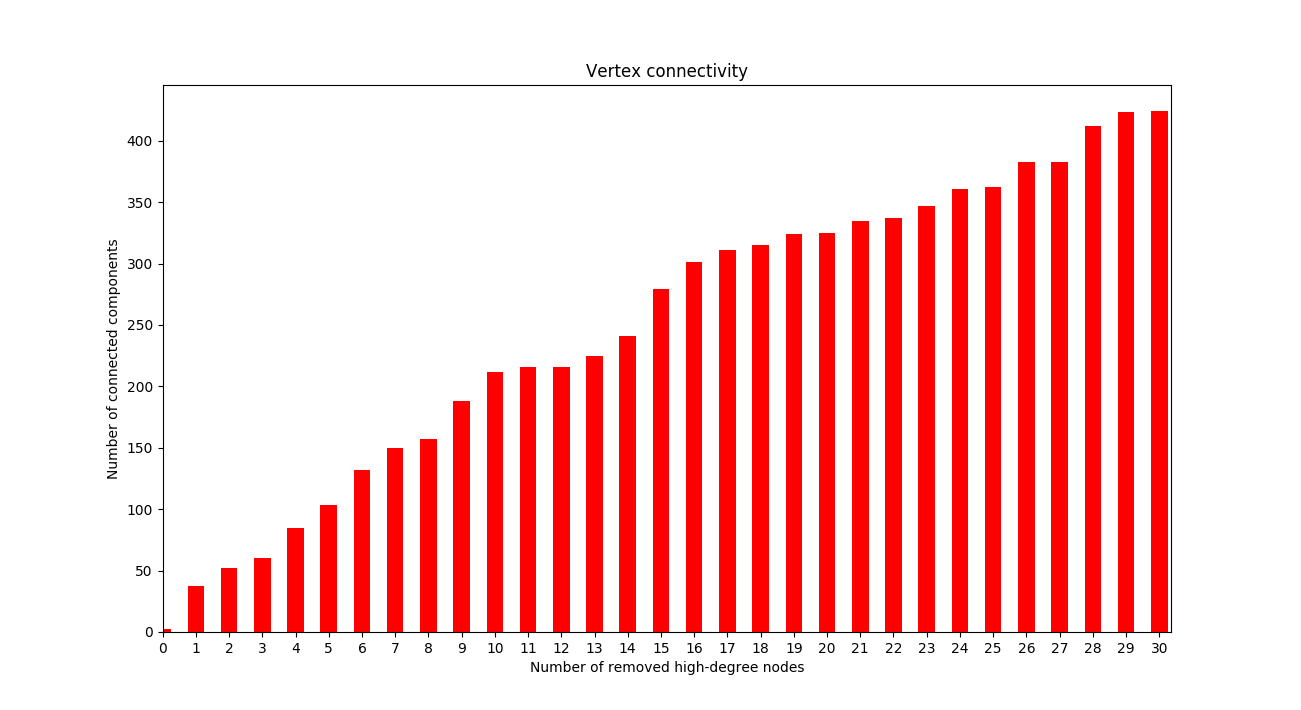}
		\captionof{figure}{LN's vertex connectivity, when all the $30$ largest hubs are removed one by one}
		\label{fig:vertexconnectivity}
	\end{minipage}%
	\begin{minipage}{.5\textwidth}
		\centering
		\includegraphics[width=\linewidth]{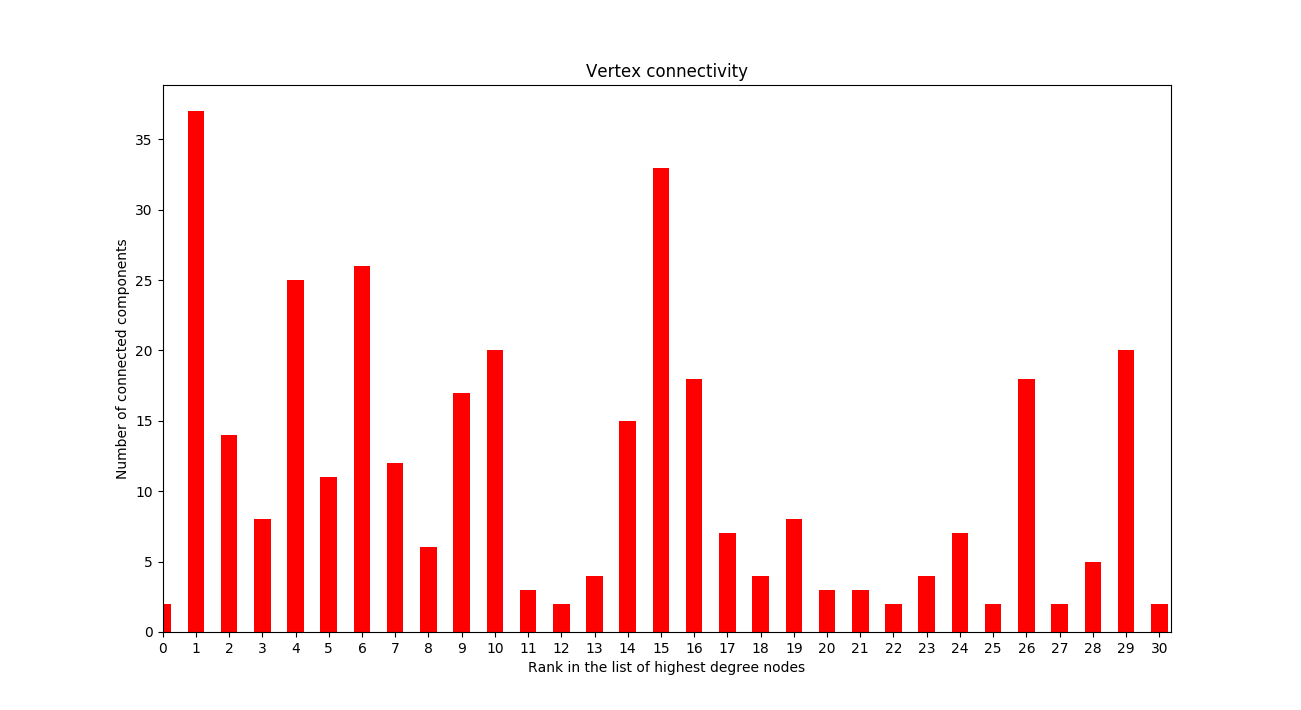}
		\captionof{figure}{LN's vertex connectivity if only one high-degree node is removed from the graph.}
		\label{fig:vertecconnectivity1removed}
	\end{minipage}
\end{figure}

Targeted attacks on LN nodes are also a major concern as the short history of LN has already shown it. On 2018 March 21st \footnote{https://www.trustnodes.com/2018/03/21/lightning-network-ddos-sends-20-nodes}, $20\%$ of nodes were down due to a Distributed Denial of Service (DDoS) attack against LN nodes. Denial of Service (DoS) attacks are also quite probable by flooding HTLCs. These attack vectors are extremely harmful, especially if they are coordinated well. One might expect that not only state-sponsored attackers will have the resources to attack a small network like LN. In the first attack scenario we removed 30 highest-degree nodes one by one starting with the most well-connected one and gradually withdraw the subsequent high-degree nodes. We recorded the number of connected components. As it is shown in Figure~\ref{fig:vertexconnectivity}. even just removing the highest-degree node\footnote{http://rompert.com/} fragments the LN graph into 37 connected components! Altogether the removal of the $30$ largest hubs incurs LN to collapse into $424$ components, although most of these are isolated vertices. This symptom can be explained by the experienced dissortativity, namely hubs tend to be at the periphery.     

We reasserted the targeted attack scenario, but for the second time we only removed one of the $30$ largest hubs and recorded the number of connected components. As it can be seen in Figure~\ref{fig:vertecconnectivity1removed} most of the hubs, 25, would leave behind several disconnected components.

\begin{wrapfigure}[19]{r}{5cm}
	\begin{tabular}{ | l | c | }
		\hline
		Network & $f_c$  \\ \hline
		Internet & $0.16$  \\ \hline
		WWW & $0.12$ \\ \hline
		Facebook & $0.28$ \\ \hline
		Euroroad & $0.59$ \\ \hline
		US Power Grid & $0.20$  \\ \hline
		Mobil Phone Call & $0.20$ \\ \hline
		Science collaboration& $0.27$ \\ \hline
		E. Coli Metabolism & $0.49$ \\ \hline
		Yeast Protein Interactions & $0.16$ \\ \hline
		LN & $0.14$ \\ \hline
	\end{tabular}
	\caption{Real networks under targeted attacks. Values of critical thresholds for other real networks are taken from \cite{barabasi2016network} and \cite{lin2017assessing}.}\label{fig:targetedfailures}
\end{wrapfigure}

Such network fragmentations are unwanted in case of LN, because they would make payment routing substantially more challenging (one needs to split the payment over several routes) or even impossible (there would be no routes at all). 

\begin{figure}[H]
	\centering
	\includegraphics[width=\linewidth]{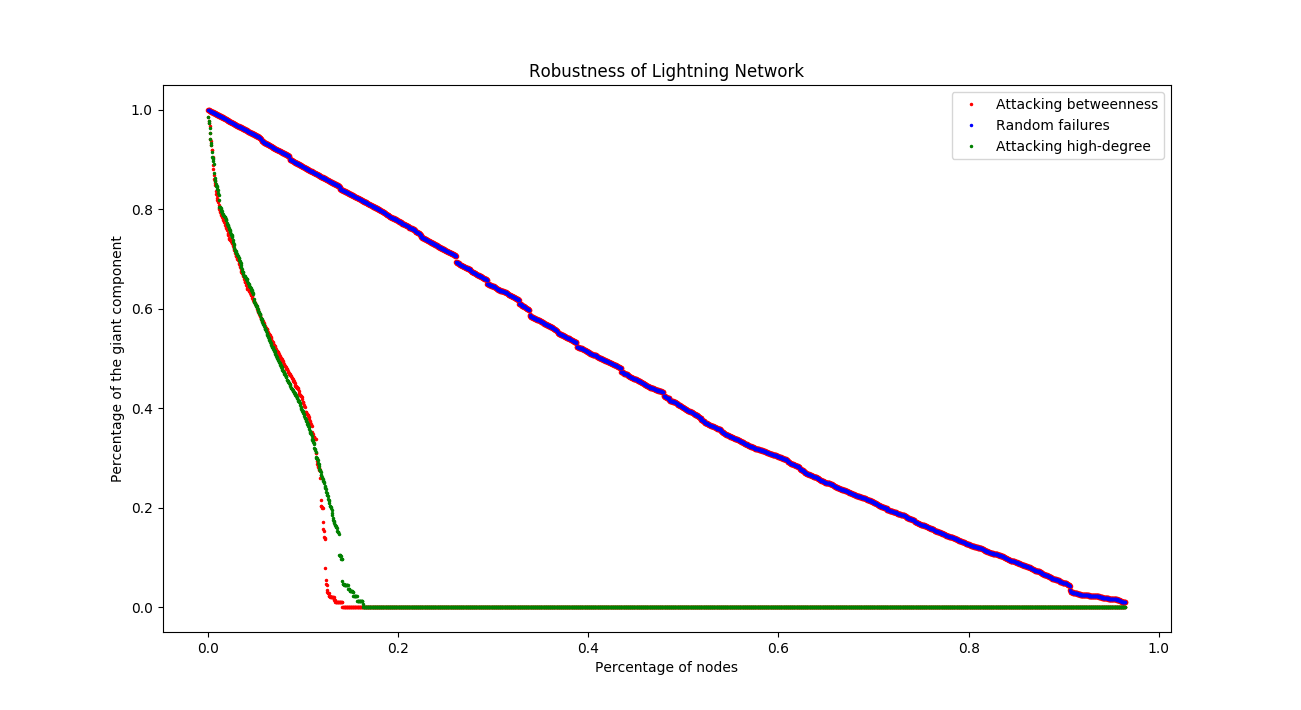}
	\captionof{figure}{Percolation thresholds for various attack scenarios: $f_{c}^{HDR}=0.1627$, $f_{c}^{HBR}=0.1409$, $f_{c}^{RND}=0.9645$}\label{fig:randhighbetween}
\end{figure}

Furthermore we estimated the percolation threshold by simulating two attacking strategies. In the first scenario we removed high degree nodes one by one (high degree removal attack, HDR) and in the second we removed nodes with the highest betweenness centrality (high betweenness removal attack, HBR). Note that in both cases after each node removal we recalculated which node has the highest degree or betweenness centrality in order to have a more powerful attack. We found out that $f_{c}=0.1627$ for removing high degree nodes, while for removing high betweenness centrality nodes $f_{c}=0.1409$, therefore choosing to remove high betweenness centrality nodes is a better strategy as it can also be seen in Figure~\ref{fig:randhighbetween}. 

\begin{figure}[h]
	\centering
	\begin{minipage}{.5\textwidth}
		\centering
		\includegraphics[width=\linewidth]{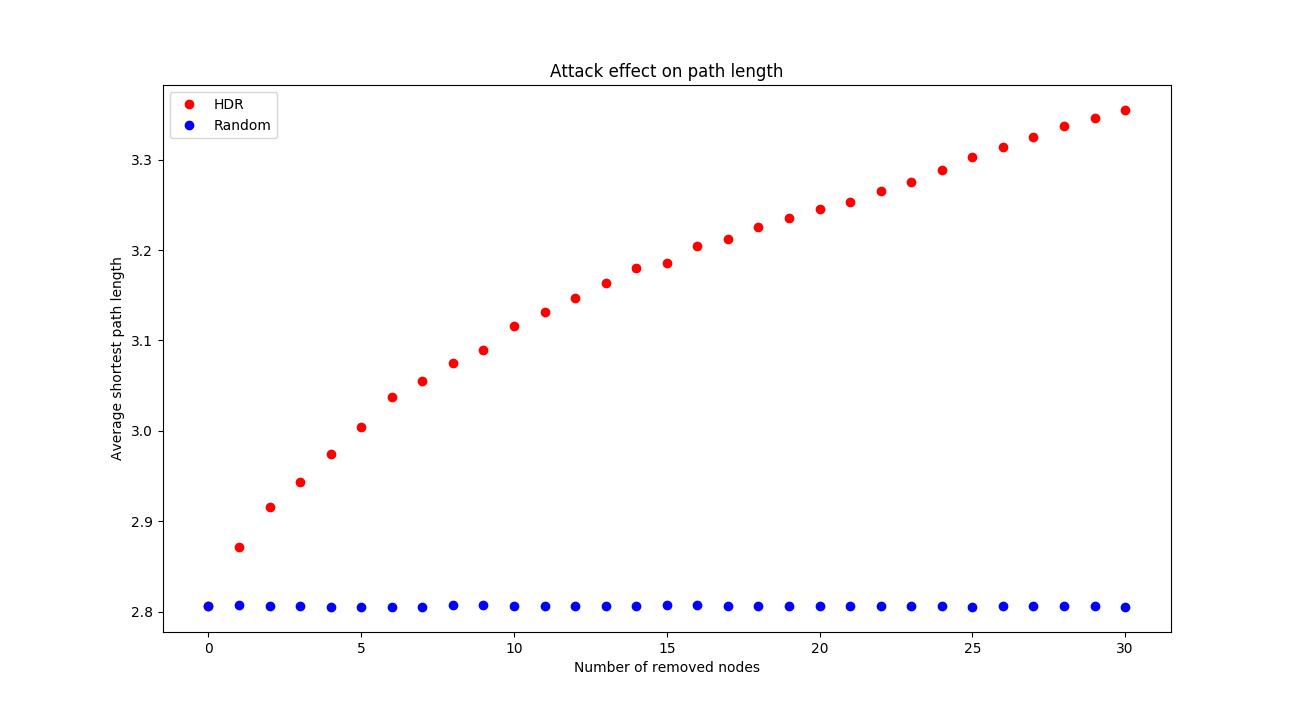}
		\captionof{figure}{High degree removal (HDR) attack effects average shortest path lengths}\label{fig:attackeffectpath}
	\end{minipage}%
	\begin{minipage}{.5\textwidth}
		\centering
		\includegraphics[width=\linewidth]{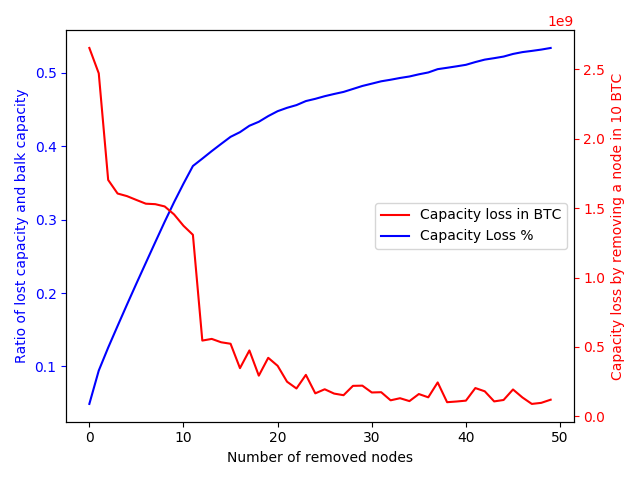}
		\captionof{figure}{Lost capacity as removing high-degree nodes}\label{fig:lostCapacity}
	\end{minipage}
\end{figure}

Node outage not only affects robustness and connectivity properties, but also affects average shortest path lengths and available liquidity. Although the outage of random nodes does not significantly increase the average shortest path lengths in LN, targeted attacks against hubs increase distances between remaining nodes. The spillage of high-degree nodes not only decreases the amount of available liquidity but also rapidly increases the necessary hops along payment routes as Figure~\ref{fig:attackeffectpath} and ~\ref{fig:lostCapacity} suggest. This could cause increased ratio of failed payments due to larger payment routes and sparser liquidity. Figure~\ref{fig:lostCapacity} demonstrates how capital allocation is centred upon a few high degree nodes, namely already the removal of as few nodes as $37$ decreases the available liquidity by more than $50\%$. Unfortunately most of these nodes are run by a handful of companies. 

\subsection{Improving LN's resilience against random failures and attacks}
Designing networks which are robust to random failures and targeted attacks appear to be a conflicting desire~\cite{barabasi2016network}. For instance a star graph, the simplest hub and spoke network, is resilient to random failures. The removal of any set of spokes does not hurt the connectedness of the main component. However it can not withstand a targeted attack against its central node, since it would leave behind isolated spokes. Furthermore when one attempts increasing robustness of a network, they do desire not to decrease the connectivity of nodes.

A similar optimization strategy of robustness and connectivity to that of~\cite{shargel2003optimization} could be applied to LN as well. We leave it for future work to empirically assess the robustness and connectivity gains if the strategy of~\cite{shargel2003optimization} would be implemented in LN client implementations. 

Nonetheless, we can still enhance the network's attack tolerance by connecting its peripheral nodes \cite{barabasi2016network}. This could be achieved by LN client implementations by implicitly mandating newcomers to connect to not only hubs, as current implementations do, but also to at least a few random nodes.

\section{Conclusion}
In summary, a better understanding of the network topology is essential
for improving the robustness of complex systems, like LN. Networks' resilience depends on their topology. LN is well approximated by the scale-free model and also its attack tolerance properties are similar to those of scale-free networks; in particular, while LN is robust against random failures, it is quite vulnerable in the face of targeted attacks. 
High-level depictions of LN's topology convey a false image of security and robustness. As we have demonstrated, LN is structurally weak against rational adversaries. Thus, to provide robust Layer-2 solutions for blockchains, such as LN and Raiden, the community needs to aim at building resilient network topologies.
\paragraph{\textbf{Acknowledgements.}}
We are grateful for the insightful comments and discussions to Chris Buckland, Daniel Goldman, Olaoluwa Osuntokun and Alex Bosworth. Furthermore we are thankful for Altangent Labs\footnote{\url{https://www.blocktap.io/}} for providing us the invaluable data.  The research at Eötvös Loránd University was partially supported by the European Union, co-financed by the European Social Fund (EFOP-3.6.2-16-2017-00012).
%
%
%
\newpage
\bibliographystyle{splncs04}
\bibliography{sample}
\end{document}